\documentclass[twocolumn]{aastex631}

\graphicspath{{./}{figures/}}

\begin{document}

\title{Uncertainty in Grain Size Estimations of Volatiles on Trans-Neptunian Objects (TNOs) and Kuiper Belt Objects (KBOs)}


\correspondingauthor{A. Emran}
\email{alemran@uark.edu}

\author[0000-0002-4420-0595]{A. Emran}
\affiliation{AR Center for Space and Planetary Sciences, University of Arkansas, Fayetteville, AR 72701, USA}

\author[0000-0002-1111-587X]{ V. F. Chevrier}
\affiliation{AR Center for Space and Planetary Sciences, University of Arkansas, Fayetteville, AR 72701, USA}

\begin{abstract}

We analyze the uncertainty in grain size estimation of pure methane (CH${}_{4}$) and nitrogen saturated with methane (\textbf{N${}_{2}$}:CH${}_{4}$) ices, the most abundant volatile materials on trans-Neptunian objects (TNOs) and Kuiper belt objects (KBOs). We compare the single scattering albedo, which determines the grain size estimation of outer solar system regolith \citep{hansen2009}, of these ices using the Mie scattering model and two other Hapke approximations \citep{hapke1993} in radiative transfer scattering models (RTM) at near-infrared (NIR) wavelengths (1 -- 5 µm). The equivalent slab (Hapke Slab) approximation model predicts results much closer to Mie scattering over the NIR wavelengths at a wide range of grain sizes. In contrast, even though the internal scattering model (ISM) predicts an approximate particle diameter close to the Mie model for particles with a 10 µm radii, it exhibits higher discrepancies in the predicted estimation for larger grain sizes (e.g., 100 and 1000 µm radii). Owing to the Rayleigh effect on single-scattering properties, neither Hapke approximate models could predict an accurate grain size estimation for the small particles (radii $\mathrm{\le}$ 5 µm). We recommend that future studies should favor the equivalent slab approximation when employing RTMs for estimating grain sizes of the vast number of TNOs and KBOs in the outer solar system.
\end{abstract}

\keywords{Classical Kuiper belt objects (250) --- Trans-Neptunian objects (1705) --- Plutoids (1268) --- Radiative transfer (1335) --- Surface ices (2117)}


\section{Introduction} \label{sec:intro}

An abundance of methane (CH${}_{4}$) and nitrogen (N${}_{2}$) ices, among others, have been detected on trans-Neptunian objects (TNOs) and Kuiper belt objects (KBOs) such as Triton \citep{cruikshank1984, cruikshank1993} and Pluto \citep{owen1993}. Eris, a dwarf planet in the Kuiper belt, also exhibits a prevalence of N${}_{2}$ and CH${}_{4}$ ices on its surface \citep{dumas2007}. Of these ices, solid CH${}_{4}$ ice has several strong absorption bands in near-infrared (NIR) wavelengths \citep{cruikshank2019}. The physical and chemical properties of these ices on TNOs and KBOs have been determined from NIR observations using the radiative transfer models (RTM) of \cite{hapke1981, hapke1993}. The RTMs enable estimation of abundances and grain size of the constituent mixtures based on albedo or reflectance of single scattered light by an average surface grain \citep{mustard2019}. For instance, \cite{protopapa2017} estimated the global scale spatial abundance and grain size distribution of ices (both volatile and non-volatile) on Pluto from Ralph/LEISA infrared spectrometer \citep{reuter2008} onboard New Horizons.

Varied results have been reported in grain size estimation of water ice at outer solar system icy bodies using different scattering models \citep{hansen2009}. Thus, it is reasonable to anticipate that the estimation of grain sizes of CH${}_{4}$ and N${}_{2}$ ices (pure and mixture) on TNOs and KBOs surfaces will have inconsistencies owing to different models used. \cite{hansen2009} argued that the differences in grain size estimation primarily arise from models used for single scattering albedo calculation rather than bidirectional reflectance models. Following the same approach, we use scattering models to NIR optical constants of CH${}_{4}$ and N${}_{2}$ ices (pure and mixture) at temperatures relevant to TNOs and KBOs surface conditions. Single scattering albedo can exactly be calculated using Mie theory \citep{mie1908} for simple geometric grains or can be estimated using Hapke approximation models \citep{hapke1981, hapke1993} from optical constants of materials. Accordingly, we analyze the single scattering albedo to estimate the inconsistencies in grain sizes using Mie theory and two approximation models from \cite{hapke1993} that have been widely used in the existing literature.

The constituent volatiles ices of TNOs and KBOs surfaces exhibit different thermodynamic phase transitions at surface conditions of the planetary bodies. For instance, the crystalline $\alpha$ (cubic) - $\beta$ (hexagonal) solid-phase transition of N${}_{2}$ occurs at 35.6K \citep{scott1976}, while CH${}_{4}$${}_{}$I - CH${}_{4}$${}_{}$II solid-phase transition occurs at $\mathrm{\sim}$21K \citep{prokhvatilov1983}. On TNOs and KBOs surfaces, the CH${}_{4}$ and N${}_{2}$ ices form solid solutions and do not coexist as pure substances \citep{cruikshank2019}. CH${}_{4}$ and N${}_{2}$ ices are completely miscible in one another and show two different solid solutions such as N${}_{2}$ diluted in CH${}_{4}$ (\textbf{CH${}_{4}$:}N${}_{2}$) and CH${}_{4}$ diluted in N${}_{2}$ (\textbf{N${}_{2}$: }CH${}_{4}$) \citep{trafton2015}. In this study, we use the NIR optical constant of pure CH${}_{4}$-I ice measured at 39K, as a proxy of methane saturated with nitrogen (\textbf{CH${}_{4}$:}N${}_{2}$), and nitrogen saturated with methane \textbf{(N${}_{2}$:}CH${}_{4}$) measured at two different temperatures of 35K and 38K. 

The rationale of using pure CH${}_{4}$ ice as the proxy \textbf{CH${}_{4}$:}N${}_{2}$ is that at a temperature of 40K, the marginal saturation limit of N${}_{2}$ in CH${}_{4}$ ice is $\mathrm{\sim}$0.035 \citep{cruikshank2019}. In the \textbf{CH${}_{4}$:}N${}_{2}$ binary system, the wavelength shift of the CH${}_{4}$ band is very small on the order of $\mathrm{\sim}$2 x 10 ${}^{-4\ }$µm \citep{protopapa2015, protopapa2017}. Using optical constants of pure CH${}_{4}$ ice as the proxy of the \textbf{CH${}_{4}$:}N${}_{2}$ system is somewhat valid below 40K \citep{protopapa2017} since the saturation limit of N${}_{2}$ in CH${}_{4}$ ice is proportional to temperature changes \citep{prokhvatilov1983}. At the 35K temperature, the crystalline $\alpha$ (cubic) phase of N${}_{2}$ is saturated with the CH${}_{4}$-I, while at the 38K, the crystalline $\beta$ (hexagonal) phase of N${}_{2}$ is saturated with the CH${}_{4}$-I${}_{\ }$ \citep{tegler2010}.

\section{Methods} \label{sec:methods}

\subsection{ Single scattering albedo}

The single scattering albedo, \textit{w}, refers to the ratio of the amount of photon scattered to the combined amount of light scattered from and absorbed by a particle:

\begin{equation} \label{Eq.1} 
w=\frac{Q_{sca}}{Q_{sca}\ +\ Q_{abs}} ,   
\end{equation} 

where $Q_{sca}$ is the scattering efficiency and $\ Q_{abs}$ is the absorption efficiency. The sum of $Q_{sca}$ and $Q_{abs}$ is termed the extinction efficiency, $Q_{ext}.$ A highly absorbing material likely exhibits a \textit{w} = 0 whereas a transparent material is more likely to show a \textit{w} = 1 \citep{shepard2017}.

In most space science applications, the \textit{w} is assumed to be the average properties, such as optical characteristics, grain size, and to some extent shape and internal structure, of particles that make up planetary regolith \citep{hapke1981}. The \textit{w} is a function of optical constants/ indices of refraction (real, \textit{n} and imaginary, \textit{k}) of particles or regolith medium \citep{mishra2021}. A particle with a larger size and moderate to larger \textit{k} tends to absorb more incident light, and therefore, exhibits a lower \textit{w} \citep{shepard2007}. A variety of approximate models has been presented by \cite{hapke1993} to calculate the \textit{w} from particle refraction indices \citep{hansen2009}. Of these models, the equivalent slab model (Hapke Slab) and internal scattering/scatterer model (ISM) are widely used in different planetary bodies \citep{li2011}. Subsequent studies \citep{hapke2001, hapke2005, hapke2012} also presented versions of ISM for surface scattering function that was originally derived from \cite{hapke1981}. We use a version of the approximate ISM and the Hapke Slab models to calculate \textit{w} from the optical constants of the ices found on TNOs and KBOs.

\subsection{Optical constants}

The optical constant (OC) of pure CH${}_{4}$ and\textbf{ N${}_{2}$:}CH${}_{4}$ were chosen based on the thermodynamics equilibrium of solid methane and nitrogen ices at different temperatures relevant to TNOs and KBOs surface conditions. As the available data permits, we use the optical constants of $\alpha$-\textbf{N${}_{2}$}:CH${}_{4}$ at 35K (1- 3.97 µm), $\beta$-\textbf{N${}_{2}$:}CH${}_{4}$ at 38K (1 - 5 µm), and CH${}_{4}$ as the proxy of \textbf{CH${}_{4}$}:N${}_{2}$ at 39K (1 - 5 µm). The optical constants of pure CH${}_{4}$ were collected from \cite{grundy2002} and the \textbf{N${}_{2}$:}CH${}_{4}$ system from \cite{quirico1992}. For detail about the optical constants used in this study see Table \ref{tab:1}. Note that the $\alpha$-\textbf{N${}_{2}$}:CH${}_{4}$ and $\beta$-\textbf{N${}_{2}$:}CH${}_{4}$ systems are the solid solutions with a concentration of $\mathrm{<}$2\% CH${}_{4}$ and the absorption coefficient is normalized to a concentration of 1 for the diluted CH${}_{4}$ in solutions (for detail refer to \cite{quirico1997}).

\begin{deluxetable*}{lcccc}
\tablenum{1}
\tablecaption{Optical constant used in this study.\label{tab:1}}
\tablewidth{0pt}
\tablehead{\colhead{Materials} & \colhead{Temperature (K)} &
\colhead{References${}^{\ a}$} & \colhead{Wavelength range (µm)} & \colhead{Notes}}
\startdata 
CH${}_{4}$ & 39 & \cite{grundy2002} & 1 - 5 & As \textbf{CH${}_{4}$}:N${}_{2}$${}_{\ }$${}^{b}$ \\
\textbf{N${}_{2}$}:CH${}_{4}$ & 35 & \cite{quirico1999}; \cite{quirico1997} & 1- 3.97 & N${}_{2}$ in $\alpha$ phase ${}^{c}$ \\
\textbf{N${}_{2}$:}CH${}_{4}$ & 38 & \cite{quirico1999}; \cite{quirico1997} & 1 - 5 & N${}_{2}$ in $\beta$ phase ${}^{d}$ \\
\enddata
\tablecomments{${}^{a}$ Optical constants are available at \href{href}{https://www.sshade.eu/}}
\noindent ${}^{b}$ Filename: optcte-Vis+NIR+MIR-CH4cr-I-39K\\
${}^{c}$ Filename: optcte-NIR-CH4-lowC-alpha-N2-35K\\
${}^{d}$ Filename: optcte-NIR-CH4-lowC-beta-N2-38K-cor
\end{deluxetable*}

\subsection{Mie calculation}

The \textit{w} for particles with simple geometrics (i.e., spherical shape) can exactly be calculated from Maxwell's equations using Mie theory \citep{mie1908} if refractive indices and particle size parameters are known \citep{moosmuller2018}. However, scattering models of highly asymmetric phase functions demand the treatment of diffraction effects from Mie \textit{w} \citep{hansen2009}. Thus, we use the \textit{$\delta$}-Eddington corrected Mie single scattering albedo\textit{w'}as compared to the estimations of \textit{w} from Hapke approximations. The \textit{$\delta$}-Eddington approximated Mie,\textit{ w' }can be calculated as \citep{wiscombe&warren1980}:

\begin{equation} \label{Eq.2} 
w'=\frac{\left(1-{\xi}^2\right)\ w}{1-{\xi}^{2\ }w}
\end{equation}

where $\xi$ is the asymmetry factor, which refers to the ratio of the forward-scattered light to the back-scattered light, calculated by Mie theory. We employ Mie \textit{w} calculation following the method described in \cite{wiscombe1979} using {\tt\string miepython} routine, a Python module licensed under the terms of the Massachusetts Institute of Technology (MIT) license. The calculated Mie \textit{w} result was then adjusted to \textit{$\delta$}-Eddington corrected Mie \textit{w'} following Eq. \ref{Eq.2}.

\subsection{Hapke Slab and ISM calculations}
The Hapke Slab approximation model is applied if the imaginary part of the optical constant k\textit{$<$ $<$}1 so that \textit{w} can be approximated as \citep{hansen2009}:

\begin{equation} \label{Eq.3} 
w=\frac{1}{\alpha DR\left(n\right)+1} 
\end{equation} 

where \textit{D} is the ``particle diameter'' and \textit{$\alpha$} is the absorption coefficient, given by:

\begin{equation} \label{Eg.4} 
\alpha =\frac{4\pi k}{\lambda } 
\end{equation} 

where \textit{$\lambda$} is the wavelength. \textit{R(n)} is the reflection function of the real part (\textit{n}) of the OC. The \textit{R(n)} can be derived as \citep{hansen2009}:

\begin{equation} \label{Eq.5} 
R\left(n\right)=\frac{1-S_e}{1-S_i} 
\end{equation} 
\\
where \textit{S${}_{e\ }$}and \textit{S${}_{i}$} are the average Fresnel reflection coefficient for externally and internally incident light, respectively \citep{hapke1993}. For the case of the slab model (\textit{k$<$ $<$1}), \textit{S${}_{e}$} can be approximated as \citep{hapke2001}:

\begin{equation} \label{Eq.6} 
S_e=\frac{{\left(n-1\right)}^2}{{\left(n+1\right)}^2}+0.05 
\end{equation} 

and the approximation for \textit{S${}_{i}$ }can be written as \citep{lucey1998}:

\begin{equation} \label{Eq.7} 
S_i=1.014-\frac{4}{n{\left(n+1\right)}^2}\  
\end{equation}

We use the ISM \citep{hapke1981} as the second approximation method where the \textit{w} can be reproduced as \citep{hapke2001}:

\begin{equation} \label{Eq.8} 
w=S_e+\left(1-S_e\right)\frac{\left(1-S_i\right)\textrm{$\Theta$}}{1-S_i\textrm{$\Theta$}}\  
\end{equation} 

\textit{$\Theta$} is the internal-transmission function of the particle is given by:

\begin{equation} \label{Eq.9} 
\textrm{$\Theta$}=\frac{r+\mathrm{exp}\mathrm{}(-\sqrt{}(\alpha \left(\alpha +s\right))\langle D\rangle }{1-r\ \mathrm{exp}\mathrm{}(-\sqrt{}\left(\alpha \left(\alpha +s\right)\right)\langle D\rangle } 
\end{equation}

where the near-surface internal scattering coefficient, \textit{s} = 0, and the internal hemispherical (diffused) reflectance, \textit{r} can be given as:

\begin{equation} \label{Eq.10} 
r=\frac{1-\ \sqrt{}\alpha \left(\alpha +s\right)}{1+\ \sqrt{}\alpha \left(\alpha +s\right)} 
\end{equation}

The average distance travel by transmitted ray i.e., mean free path of photon \textit{$\langle $D$\rangle $} as a function of \textit{n} and particle diameter, \textit{D} for perfectly spherical particle can be written as \citep{hapke2005}:

\begin{equation} \label{Eq.11} 
\langle D\rangle \ =\frac{2}{3}\left(n^2-\frac{1}{n}{\left(n^2-1\right)}^{\frac{3}{2}}\right)D 
\end{equation} 

In the ISM model, \textit{S${}_{i}$} is derived from Eq. \ref{Eq.7} while \textit{S${}_{e}$}${}_{}$is a function of both \textit{n} and \textit{k}, as given by:

\begin{equation} \label{Eq.12} 
S_e=\frac{{\left(n-1\right)}^2+k^2}{{\left(n+1\right)}^2+k^2}+0.05 
\end{equation}

\section{Results}

\subsection{Calculated single scattering albedo}

\begin{figure*}[ht!]
\plotone{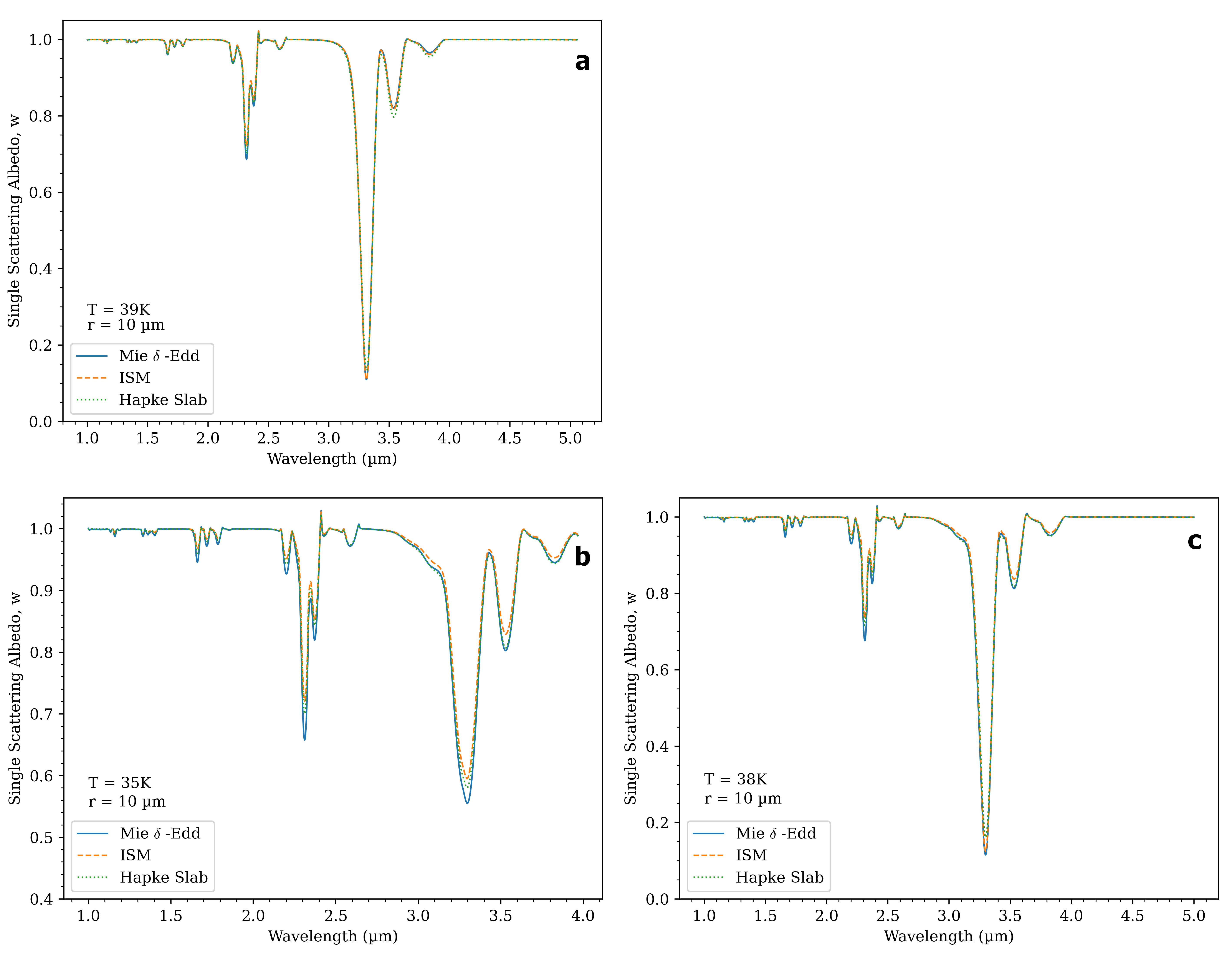}
\caption{Plots for single scattering albedo of pure CH${}_{4}$ at 39K (a), $\alpha$-\textbf{N${}_{2}$}:CH${}_{4}$ at 35K (b), and $\beta$-\textbf{N${}_{2}$}:CH${}_{4}$ at 38K (c) for a grain 10 µm radius. The dashed solid line presents \textit{$\delta$}-Eddington corrected Mie, dashed yellow represents ISM, and the dotted green represents Hapke Slab results. The plots were smoothened using the \textit{Savitzky-Golay} filter \citep{savitzky1964}.
\label{Fig1}}
\end{figure*}

To evaluate how the Mie and Hapke models produce different \textit{w} at a specific particle diameter, we compare \textit{w} of a 10 µm radius particle for pure CH${}_{4}$ at 39K, $\alpha$-\textbf{N${}_{2}$}:CH${}_{4}$ at 35K, and $\beta$-\textbf{N${}_{2}$}:CH${}_{4}$ at 38K (Fig. \ref{Fig1}). The calculated \textit{w} exhibits small spikes over the wavelengths, and therefore we smoothen our calculated \textit{w} curves by applying the \textit{Savitzky-Golay} filters \citep{savitzky1964}, an algorithm typically used for signal processing, with a 3${}^{rd}$ order of the polynomial fit.

The result of \textit{w} from Hapke Slab and ISM calculated for pure CH${}_{4\ }$ice shows a varying degree of closeness to the result of Mie calculation at different NIR wavelengths (Fig \ref{Fig1}a). More specifically, at shorter wavelengths up to 2 µm, both ISM and slab model exactly follow and mimic the results of Mie calculation. However, at wavelengths of 2.13 and 2.37 µm, the slab model results show slightly higher \textit{w} than the Mie calculation. This indicates that the Hapke Slab model predicts a slightly smaller pure CH${}_{4}$ grain size compared to the Mie model at these wavelengths \citep{hansen2009}. The ISM \textit{w} is closer to the Mie result at longer wavelengths at 3.3 and 3.5 µm -- indicating a similar grain-size prediction by ISM to the Mie at these wavelengths. In contrast, the Hapke Slab predicts a smaller and larger pure CH${}_{4}$ ice grain compared to the Mie results at 3.3 and 3.5 µm, respectively. At 3.82 µm, both approximation models predicate a slightly higher grain size than the Mie results.

The \textbf{N${}_{2}$}:CH${}_{4}$ system results (Fig. \ref{Fig1}b and \ref{Fig1}c) show that the Hapke Slab model, overall, produces \textit{w} values much closer to the Mie model than ISM at both 35 and 38K temperatures, except at $\mathrm{\sim}$ 3.3 µm wavelength for 38K where the ISM's \textit{w} gets closer to Mie. In the $\alpha$-\textbf{N${}_{2}$}:CH${}_{4}$ ice (Fig. \ref{Fig1}b), the \textit{w} calculated from the Hapke Slab model is much closer to the Mie model while ISM predicted a much higher \textit{w} than Mie calculations over the entire NIR wavelengths. This indicates a much smaller $\alpha$-\textbf{N${}_{2}$}:CH${}_{4}$ ice grain-size prediction by ISM. Likewise, in the case of the $\beta$ -\textbf{N${}_{2}$}:CH${}_{4}$ system (Fig. \ref{Fig1}c), the slab model result follows much closer to the Mie result over the wavelengths, except for the 3.3 µm where ISM gets closer to Mie than slab result. This implies that the Hapke Slab model produces a slightly smaller grain size around 2.3, 3.3, and 3.5µm at 35K and 2.3 and 3.5 µm at 38K than the Mie model. Around the wavelength of 3.3µm at 38K, the ISM predicts a slightly smaller grain size than the Mie compared to Hapke Slab to Mie. Overall, for the \textbf{N${}_{2}$}:CH${}_{4}$ systems, ISM predicts a much smaller grain size than the Mie compared to the slab model to Mie over the NIR wavelengths for the particles with a 10 µm radii.

\subsection{Discrepancies in grain size estimation}

\begin{figure*}[ht!]
\plotone{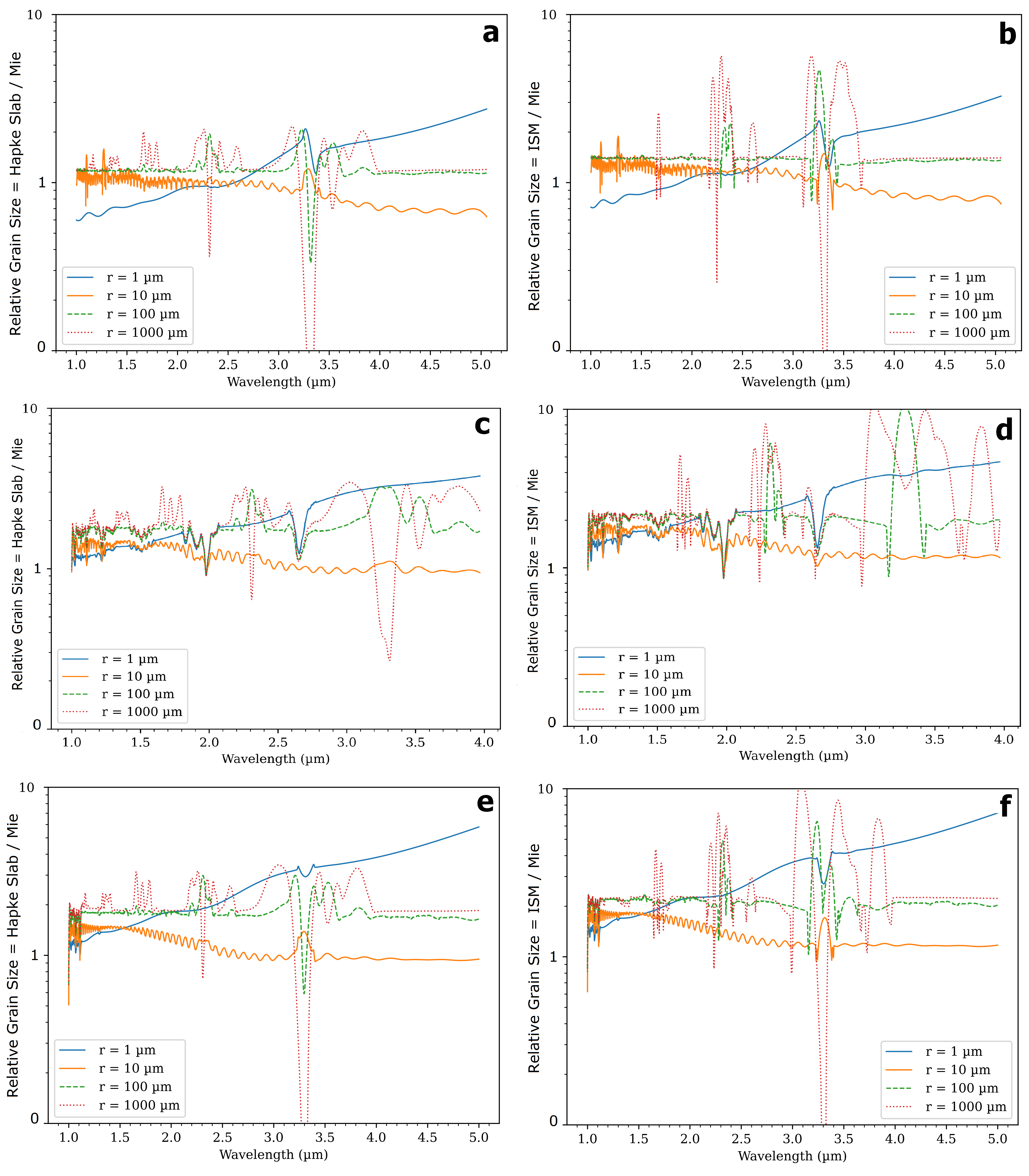}
\caption{Grain sizes determined from the Hapke Slab (left column) and ISM (right column) using spectra calculated using the Mie model at different particle radii of 1, 10, 100, 1000 µm for pure CH${}_{4}$ ice at 39K (top row), $\alpha$-\textbf{N${}_{2}$}:CH${}_{4}$ at 35K (middle row), and $\beta$-\textbf{N${}_{2}$}:CH${}_{4}$ at 38K (bottom row). The resulting grain sizes are normalized to the input grain sizes. The plots were smoothened using the \textit{Savitzky-Golay} filter \citep{savitzky1964}.
\label{Fig2}}
\end{figure*}

We compare the relative grain size predicted by the approximated Hapke models to the Mie model over the NIR wavelengths. To end that, we first estimate the \textit{$\delta$}-Eddington corrected Mie \textit{w'} at grain radii of 1, 10, 100, and 1000 µm. Then we estimate the grain sizes corresponding to these Mie \textit{w'} values by applying the inverse ISM and Hapke Slab models. The \textit{w} is a non-linear function of diameter (\textit{D}) in both Hapke approximation models (Eq. \ref{Eq.3} and \ref{Eq.8}). We solve the non-linear equations of Hapke Slab and ISM for \textit{D} using Powell's hybrid (dogleg) method \citep{powell1970, chen1981}. Lastly, we estimate the relative discrepancies in grain size determination by normalizing the estimated grain sizes from the Hapke Slab and ISM models to the Mie grain size for pure CH${}_{4}$, $\alpha$-\textbf{N${}_{2}$}:CH${}_{4}$, and $\beta$-\textbf{N${}_{2}$}:CH${}_{4}$ (Fig. \ref{Fig2}).

For pure CH${}_{4}$ ice, the Hapke Slab model (Fig. \ref{Fig2}a) better predicts the grain sizes than the ISM model (Fig. \ref{Fig2}b). Overall, the predicted grain sizes by the Hapke Slab model are within $\mathrm{\sim}$ 20\% of the grain size of Mie, whereas the discrepancies in the predicted grain sizes by the ISM model are much higher. However, around the 3.3 µm region, neither model did predict grain sizes very well to Mie's results. Even though the slab model can predict twice as much as the Mie results at some wavelengths, for instance, at around 2.3$\mathrm{\sim}$2.4 µm region for larger grain sizes (e.g., 1000 µm radii), the ISM predicated results still show much more discrepancies than slab model results at these wavelengths. The continuous rise of the 1 µm curve is due to largely the Rayleigh effect \citep{hansen2009} from the Mie model that the Hapke Slab method does not model. Likewise, the Rayleigh effect on single-scattering properties is not modeled by the ISM. Both Hake Slab and ISM provide a better prediction at 10 µm grain size, within $\mathrm{\sim}$30\% of the grain size of Mie. However, larger grain-sized were not modeled very well by the ISM and predicted the grain sizes that are many times the grain size of Mie (Fig. \ref{Fig2}b). 

The \textbf{N${}_{2}$}:CH${}_{4}$ system results (Fig. \ref{Fig2}c, d for 35K and Fig. \ref{Fig2}e, f for 38K) show that, overall, the Hapke Slab model has better a prediction, and thus lower discrepancies in grain size estimation to the Mie model than the ISM model. Owing to the Rayleigh effect on single-scattering properties modeled by Mie theory \citep{hansen2009} but not by either of the approximation models for smaller grain sizes (e.g., 1 µm), there is a trend of continuously increasing discrepancies in grain sizes estimation by both Hapke Slab and ISM. Similar to the case of CH${}_{4}$ ice, both Hapke Slab and ISM models fit best at 10 µm, where it is within $\mathrm{\sim}$30\% of the Mie result, for the $\alpha$-\textbf{N${}_{2}$}:CH${}_{4}$ and $\beta$-\textbf{N${}_{2}$}:CH${}_{4}$ ices. For larger grain radii (e.g., 100 and 1000 µm) the Hapke Slab exhibits comparatively lower discrepancies than ISM results and predicted grain sizes within the twice of the Mie over most of the NIR wavelengths, except at 3.3µm for both temperatures. In contrast, though ISM models a good fit at 10 µm, the larger particles (e.g., 100 and 1000 µm) exhibit higher degrees of discrepancies compared to the Hapke Slab results to Mie. This characteristic result from the \textbf{N${}_{2}$}:CH${}_{4}$ system is also consistent with the result for water ice grain sizes of Enceladus \citep{hansen2009}.

\subsection{Effect of the absorption coefficient}
\begin{figure*}[ht!]
\plotone{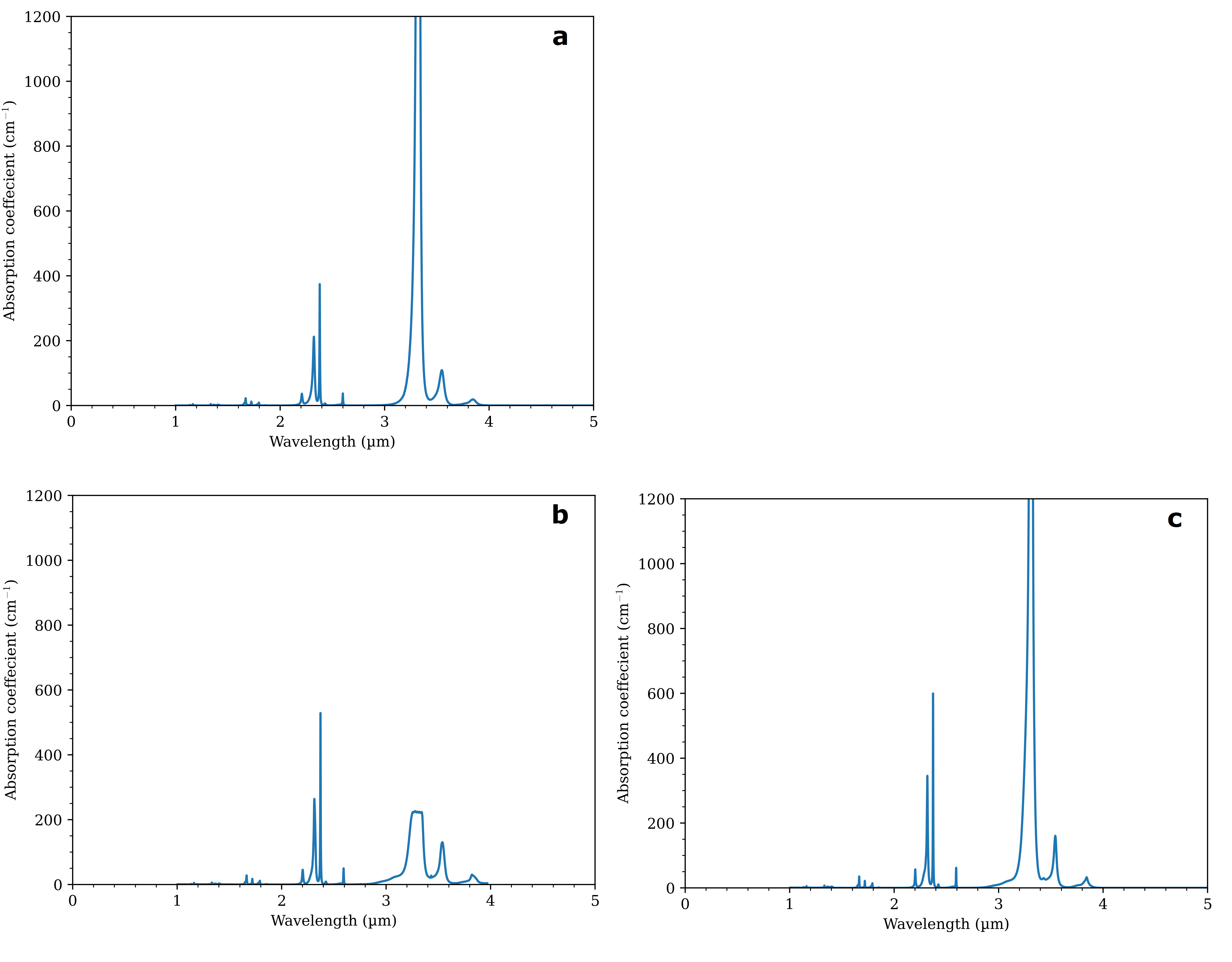}
\caption{Absorption coefficient of pure CH${}_{4}$ ice at 39K (a), CH${}_{4}$ saturated with $\alpha$-N${}_{2}$ ice at 35K (b), and CH${}_{4}$ saturated with $\beta$- N${}_{2}$ ice at 38K (c). Pure CH${}_{4}$ has an absorption coefficient peak at 3.3 µm. The absorption coefficient is much higher over the wavelength around 3.3 µm at 38K compared to 35K for CH${}_{4}$saturated with N${}_{2}$ ices.
\label{Fig3}}
\end{figure*}

The distribution of absorption coefficients (\textit{$\alpha$}) of pure CH${}_{4}$ and CH${}_{4}$ saturated with N${}_{2}$ ices over the NIR wavelength region is given in Fig. \ref{Fig3}. Absorption coefficients of pure CH${}_{4}$ ${}_{}$ show a peak around the 3.3 µm wavelength (Fig. \ref{Fig3}a). Though absorption coefficients are quite similar for \textbf{N${}_{2}$:}CH${}_{4}$ ${}_{}$ systems both 35 and 38K, at the latter temperature (Fig. \ref{Fig3}c), absorption coefficients at around 3.3 µm wavelength are much higher than the former temperature (Fig. \ref{Fig3}b). Pure CH${}_{4}$ and $\beta$-\textbf{N${}_{2}$:}CH${}_{4}$ ${}_{}$ice show a similar higher absorption coefficient at 3.3 µm wavelength. The anomalies in grain size prediction of larger particles by the Hapke Slab method for pure CH${}_{4}$ and $\beta$-\textbf{N${}_{2}$}:CH${}_{4}$ at 3.3 µm are, perhaps, due to the larger \textit{k} value at this NIR wavelength region.

The absorption coefficient peak at 3.3 µm at 38K and 39K are consistent with the fact of lower single scattering albedo or higher absorption. This is also evident in the relative grain-size curve for 10 µm, the best grain size prediction by the Hapke approximated models. The grain-size curves at 38K (Fig. \ref{Fig2}a, b) and 39K (Fig. \ref{Fig2}e, f) follow a continuum over the NIR wavelengths except for a ``dome'' in the 3.3 µm region -- indicating a predicted larger grain size corresponding to higher absorption. We compare \textit{$\alpha$} from our result to that of water (H${}_{2}$O) ices in outer solar system bodies as given in Fig. 4 of \cite{hansen2009}. Water ice has a higher absorption coefficient in most NIR wavelength regions compared to both pure CH${}_{4}$ and \textbf{N${}_{2}$:}CH${}_{4}$ systems. This implies that pure CH${}_{4}$ and \textbf{N${}_{2}$:}CH${}_{4}$ ice grains have higher reflectance and lower absorption compared to water ice grains over the NIR wavelengths. One possible interpretation of this comparison is that grain size estimation of H${}_{2}$O ice using the Hapke approximation models may predict relatively larger grains compared to pure CH${}_{4}$ and \textbf{N${}_{2}$:}CH${}_{4}$ ices grains at the outer solar systems bodies. However, this interpretation is based on the distribution of absorption coefficient over NIR wavelengths, while other factors are involved, different interpretations are also plausible.

\subsection{The interplay between albedo, absorption coefficient, and grain size}

\begin{figure*}[ht!]
\plotone{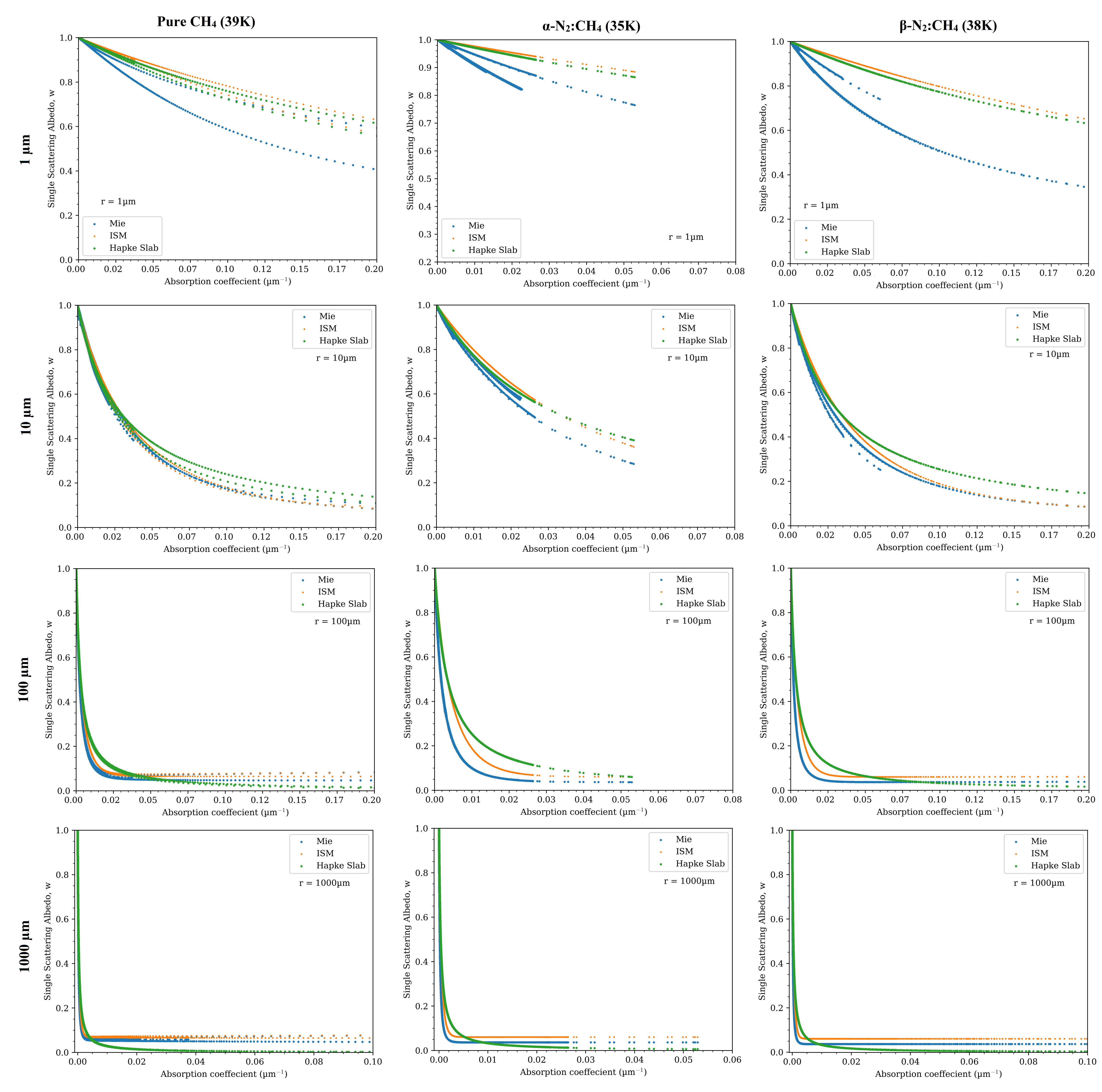}
\caption{Plot of the single scattering albedo against absorption coefficient for pure CH${}_{4}$ ice particle at 39K (left column), $\alpha$ -\textbf{N${}_{2}$}:CH${}_{4}$ particle at 35K (middle column), and $\beta$ -\textbf{N${}_{2}$}:CH${}_{4}$ particle at 38K (right column). The subplots show the distribution of \textit{w} at grain radii at 1 µm (upper row), 10 µm (2${}^{nd}$ row), 100 µm (3${}^{rd}$ row), and 1000 µm (bottom row). In all subplots, the scattering albedo from the Mie calculation (blue), ISM (yellow), and Hapke Slab (green). The Mie calculation follows two different paths due to the Rayleigh scattering effect at lower grain-size particles.
\label{Fig4}}
\end{figure*}

We analyze characteristic distribution \textit{w }as a function of \textit{$\alpha$} for Mie, ISM, and Hapke Slab models. The \textit{w }verses \textit{$\alpha$} for pure CH${}_{4}$ (left column), $\alpha$-\textbf{N${}_{2}$}:CH${}_{4}$ (middle column), and $\beta$-\textbf{N${}_{2}$}:CH${}_{4}$ (right column), ices at different grain radii (in rows) are given in Fig. \ref{Fig4}. In 1 µm gain size graphs (upper row of Fig. \ref{Fig4}), the Mie calculation follows two different paths, and their separation is largely due to the Rayleigh scattering effect at smaller grain-sized particles \citep{hansen2009}. At all temperatures, each approximation model follows a relatively linear path while one of the Mie paths shows a slightly exponential fall of albedo for pure CH${}_{4}$ and $\beta$-\textbf{N${}_{2}$}:CH${}_{4}$ ices plots. In the \textbf{N${}_{2}$}:CH${}_{4}$ systems, the Hapke Slab and ISM models follow a relatively similar path. However, one of the routes of Mie plots follows closely to the Hapke Slab and ISM plots for pure CH${}_{4}$. Most of the weakly absorbing points with higher single-scattering albedos are located below the absorption coefficient of $\mathrm{\sim}$ 0.07 µm${}^{-1}$ in pure CH${}_{4}$ and $\beta$-\textbf{N${}_{2}$}:CH${}_{4}$ ices, whereas below 0.03 µm${}^{-1}$ in $\alpha$-\textbf{N${}_{2}$}:CH${}_{4}$ ice.

There is an exponential drop in the single scattering plots for 10 µm graphs at all temperatures and thermodynamics ice phases of methane and nitrogen (2${}^{nd}$ row of Fig. \ref{Fig4}). The Rayleigh scattering effect (i.e., two separate routes of Mie plots) is also evident here, but in the weakly separated routes. The plots of Hapke Slab and ISM approximate models follow along (similar trend) one of the Mie paths for all thermodynamic pure and mixture ice phases --indicating a close fit of these approximates models to Mie result at this grain size. There are few points beyond the absorption coefficient of 0.07 µm${}^{-1}$, that we consider roughly the exponential breakpoint, for pure CH${}_{4}$ and $\beta$-\textbf{N${}_{2}$}:CH${}_{4}$, whereas this breakpoint is $\mathrm{\sim}$0.03 µm${}^{-1}$ for $\alpha$-\textbf{N${}_{2}$}:CH${}_{4}$. 

For the 100 µm plots (3${}^{rd}$ row of Fig. \ref{Fig4}), there is a steep decline of the plots for all phases. Neither of the approximation models consistently follow the Mie route; at some absorption coefficient values, ISM gets closer to Mie while at other points Hapke Slab model closely follows the Mie path. The breakpoint of the steeply declined plots is roughly around 0.01 µm${}^{-1}$ and most of the absorption points reside below this threshold where single scattering albedo value varies in a wide range (e.g., \textit{w} = 0.1 -- 1.0). In the 1000 µm plots (lower row of Fig. \ref{Fig4}), the single scattering albedo plots more steeply decline at all thermodynamics ice phases of methane and nitrogen. However, the breakpoint of the albedo plots is $\mathrm{<}$ 0.01 µm${}^{-1}$, meaning the single scattering albedo steeply falls at a lower absorption coefficient for larger grain sizes. Similar to the 100 µm plots, in 1000 µm plots neither of the approximation models consistently follow the Mie route.

\section{Discussion and conclusion}

Amidst inconsistent results in the grain size estimation of water ice on outer solar system bodies \citep{hansen2009}, we analyze the relative differences of grain size estimation for pure CH${}_{4}$ and \textbf{N${}_{2}$:}CH${}_{4}$ ices relevant to TNOs and KBOs. We calculate the single scattering albedo using Mie and two other Hapke approximations models for these ices at NIR wavelengths. Compared to ISM, the Hapke Slab approximation model predicts much closer results to Mie scattering results. In pure CH${}_{4}$ ice, the overall estimated grain size differences between the Hapke Slab and ISM are about 10\% for the particles with different grain radii, except in the case of larger particles and longer wavelengths where the imaginary part of the refractive index is much larger. Similarly, in \textbf{N${}_{2}$:}CH${}_{4}$ systems, the average differences between the estimated grain sizes from Hapke Slab and ISM are around $\mathrm{\sim}$20\% for wide ranges of grain size. 

Both Hapke Slab and ISM were found to be appropriate models for a grain size radius of 10 µm. For smaller grain-sized particles (radii of $\mathrm{\le}$5 µm), neither approximate model predicts an accurate grain size due to the Rayleigh effect. For larger grains at longer wavelengths, particularly at wavelengths with higher absorption coefficient values, the ISM predicted grain sizes exhibit larger anomalies compared to the Mie result. Overall, the results estimated prediction at different grain sizes indicate that the Hapke Slab model is the more well-predicted model to the Mie result over the NIR optical constants of pure CH${}_{4}$ and \textbf{N${}_{2}$:}CH${}_{4}$ systems while ISM's predictions show higher discrepancies. Existing literature indicates that the discrepancy in grain-size determination for larger particles at longer wavelength using the ISM can be mitigated by fine-tuning the value of free parameters. For instance, \citet{roush2007} used an increased value of \textit{s} in their Eq. 2 and 3 from 10${}^{-17}$ to 1.25 cm${}^{-1}$ to fit the modeled spectrum of gypsum power to measured spectrum \citep{hansen2009}. 

The particle diameter in the Hapke Slab model is associated with a scaling factor that varies from 3/4 to 4/3 \citep{hansen2009}. In ISM, similar uncertainty in selecting the mean free path of photon \textit{$\langle $D$\rangle $ }by using different scale factors to effective grain size \textit{D}. For instance, for spherical particles, \textit{$\langle $D$\rangle $} can be approximated to $\mathrm{\cong }$ 0.9\textit{D} \citep{hapke2012}, 2\textit{D}/3 \citep{melamed1963}, etc., while for irregular particles \textit{$\langle $D$\rangle $ }= 0.2D \citep{shkuratov2005}. This study uses a scale factor of 1 in the Hapke Slab model. We use the mean free path of photon \textit{$\langle $D$\rangle$} calculated from effective grain size (\textit{D}) and the real part (\textit{n}) of the refractive index using Eq. \ref{Eq.12}. The Hapke Slab and IMS can, therefore, be improved by fine-tuning the scale factors to the approximate models. The slab model accounts only \textit{n} while the ISM considers both the \textit{n} and \textit{k} parts of the refractive indices in the calculation of the average Fresnel reflection coefficients. If the internal scattering coefficient, \textit{s} is set to 0, the internal hemispherical (diffused) reflectance, \textit{r }equals 0. However, the relationship between the internal scattering coefficient (\textit{s}) and effective particle diameter (\textit{D}) has also been expressed as \textit{s} = 1/\textit{D } \citep{sharkey2019}. This relationship also defines that the number of scattering events within a single grain is set to 1. Consequently, this relationship indicates that the value of \textit{s} cannot be 0.

The application of the scattering (and absorption) properties of Mie spheres has been shown to be satisfactory for varied non-spherical particle shapes \citep{grenfell1999, neshyba2003, grenfell2005}. The Mie formulation accurately predicts the scattering properties of equivalent spheres of particles, even it can produce satisfactory scattering results for irregular particles \citep{neshyba2003}. Thus, the size of spherical particles from the Mie model can somewhat be analogous to non-spherical particles \citep{hansen2009}. Moreover, the Mie scattering formulation properly accounts for the Rayleigh effects of scattering properties for smaller particle sizes that are ignored by the Hapke approximation models. Therefore, based on our results, we recommend using the Mie calculation for radiative transfer modeling to unknown spectra of TNOs and KBOs. Our results show that the Hapke Slab approximation model, overall, well predicts the grain size to the Mie model over the NIR wavelengths. Thus, if the Hapke approximation models are to choose, we suggest using the equivalent slab model over the internal scattering model in estimating the pure CH${}_{4}$ and \textbf{N${}_{2}$:}CH${}_{4}$ ice grain sizes on trans-Neptunian objects and Kuiper belt objects. Our study provides a guideline for the future application of RTM in estimating the ice grain sizes at TNOs and KBOs.

\begin{acknowledgments}
The authors would like to thank anonymous reviewers for their useful comments.
\end{acknowledgments}

\begin{appendix}
\section{Notation}
List of notation and symbols used in this paper

\textit{$\langle $D$\rangle$}  mean free path of a photon 

\textit{D} particle diameter

\textit{$\xi$} asymmetry parameters of Mie theory

\textit{k} imaginary part of the refractive index

\textit{n} real part of the refractive index

\textit{r }internal diffused reflectance

\textit{R(n) }reflection function

\textit{s }internal scattering coefficient

\textit{S${}_{e}$ }Fresnel reflection coefficient for externally incident light,\textit{}

\textit{S${}_{i}$ }Fresnel reflection coefficient for internally incident light,

\textit{w} single scattering albedo

\textit{w' $\delta$}-Eddington Mie single scattering albedo\textit{ }

\textit{$\alpha$} absorption coefficient

\textit{$\lambda$} wavelength

\textrm{$\Theta$} internal transmission coefficient

\end{appendix}


\bibliography{sample631}{}
\bibliographystyle{aasjournal}



\end{document}